\documentclass[lettersize,journal]{IEEEtran}
\usepackage{amsmath,amsfonts}
\usepackage{algorithmic}
\usepackage{algorithm}
\usepackage{array}
\usepackage[caption=false,font=normalsize,labelfont=sf,textfont=sf]{subfig}
\usepackage{textcomp}
\usepackage{stfloats}
\usepackage{url}
\usepackage{verbatim}
\usepackage{graphicx}
\usepackage{cite}

\begin{document}

\title{Reconfigurable Intelligent Surfaces: Performance Assessment Through a System-Level Simulator}

\author{\IEEEauthorblockN{Björn Sihlbom\IEEEauthorrefmark{1},
		Marios~I.~Poulakis\IEEEauthorrefmark{1},
		and Marco~Di~Renzo\IEEEauthorrefmark{2} \\
		\IEEEauthorblockA{\IEEEauthorrefmark{1}Huawei Technologies Sweden AB, Sweden} \\
		\IEEEauthorblockA{\IEEEauthorrefmark{2}Paris-Saclay University -- CNRS and CentraleSupelec, France} \\
		Email: \{bjorn.sihlbom, marios.poulakis\}@huawei.com, marco.di-renzo@universite-paris-saclay.fr }
%         <-this % stops a space
%\thanks{This paper was produced by the IEEE Publication Technology Group. They are in Piscataway, NJ.}% <-this % stops a space
%\thanks{Manuscript received April 19, 2021; revised August 16, 2021.}
}

% The paper headers
%\markboth{Journal of \LaTeX\ Class Files,~Vol.~XX, No.~X, October~2021}%
%{Shell \MakeLowercase{\textit{et al.}}: A Sample Article Using IEEEtran.cls for IEEE Journals}

%\IEEEpubid{0000--0000/00\$00.00~\copyright~2021 IEEE}
% Remember, if you use this you must call \IEEEpubidadjcol in the second
% column for its text to clear the IEEEpubid mark.

\maketitle

\begin{abstract}
Reconfigurable intelligent surfaces (RISs) are considered a promising technology for boosting the coverage and for enhancing the spectral efficiency of wireless systems, as well as for taming the wireless environment. The potential benefits of RISs are currently being analyzed and various approaches are being proposed to address some challenges for their integration in wireless networks. Currently available studies to quantify the potential gains of deploying RISs in wireless networks are limited to simple network topologies, while no system-level assessments have been reported to date. Network-level, e.g., on the scale of hundreds of square meters, simulations are, however, the first step to quantify the actual value of emerging technologies and the steppingstone before considering large scale system-level experimental evaluations and network deployments. Towards this direction, this article reports the first system-level simulation results and analysis of an RIS deployment in a typical urban city that is served by a fifth-generation cellular network. The obtained system-level simulation results unveil that the benefits of RISs vary depending on the operating frequency and the size of the surfaces. Specifically, we investigate the performance improvement that RISs can provide, in terms of outdoor and indoor coverage and ergodic rate, when deployed in mid (C-band) and high (millimeter-wave) frequency bands. For example, the obtained results unveil that the deployment of RISs enhances the coverage of cell-edge users from 77\% to 95\% in the C-band and from 46\% to 95\% in the millimeter-wave band.
\end{abstract}


\section{Introduction}
\IEEEPARstart{R}{econfigurable} intelligent surfaces (RISs) constitute an emerging transmission technology that is under evaluation for possible integration in future programmable wireless networks. For example, the European Telecommunications Standards Institute (ETSI) has recently launched a focused industry specification group whose objective is to coordinate current research and development efforts on RISs for facilitating the potential adoption of this technology in future telecommunication standards\footnote{https://www.etsi.org/committee/1966-ris.}. 

An RIS is an engineered (intelligent) surface, which is made of electromagnetic metamaterials and is equipped with simple electronic circuits (e.g., diodes or tunable capacitors), that is capable of intentionally and smartly controlling the propagation of the electromagnetic waves in complex wireless propagation environments, so as to boost the signal quality in a cost-effective and energy-efficient manner \cite{JSAC_2020}. The potential benefits of RISs in wireless networks depend on the carrier frequency. In particular, an RIS deployed in the mid frequency spectrum, e.g., the C-band, of fifth-generation (5G) cellular networks is expected to primarily enhance the capacity of cell-edge users, to improve the channel rank of multiple-input multiple-output (MIMO) transmission links, and to suppress the network interference. An RIS deployed in the high frequency spectrum, e.g., the millimeter-wave (mmWave) band, is expected to primarily enhance the signal-to-noise-ratio (SNR) in weak coverage areas and blind zones, by overcoming non-line-of-sight propagation conditions and by enhancing outdoor-to-indoor communications. In this context, an RIS may constitute a cost-effective, smart, reconfigurable, and easy to deploy alternative to relays, which is able to operate in full-duplex mode without self-interference \cite{OpenJournal_2019}.

While the potential advantages and gains of deploying RISs in wireless networks are under discussion and analysis, it is acknowledged that several  important design challenges need to be tackled for possibly integrating RISs in future telecommunication standards. These include the definition of electromagnetically consistent models for metasurfaces, as well as the derivation of channel models for RIS-aided links and their integration into ray tracing simulators \cite{PIEEE-CommunModels_2021}, the development of scalable algorithms for optimizing the operation, the placement and deployment of RISs \cite{Rui-TutorialTCOM_2021}, the design of efficient channel estimation algorithms, and feedback channels for controlling and configuring RISs  \cite{Cunhua-TwoTimeScale_2021}, as well as the ownership and interference management in multi-operator networks. In addition, the identification of the most suitable use cases and deployment scenarios for RISs is currently being discussed. Example of use cases that were recently showcased at the Docomo Open House 2021 include the deployment of transparent RISs for enabling outdoor-to-indoor communications in the mmWave frequency band\footnote{https://docomo-openhouse.jp/2021/en/exhibition/022/.} by using, e.g., transparent metamaterial films without spoiling the appearance of windows\footnote{https://www.sekisui.co.jp/electronics/en/application/film.html.}.

Testbed validations, real-world experiments, and field trials are essential in order to demonstrate, in existing 5G network deployments, the actual value of emerging technologies like RISs. In the recent open technical literature, various testbed platforms have been reported. Examples of hardware prototypes include two-state digital surfaces for anomalous reflection in the sub-6 GHz \cite{Wankai-Sub6_2021} and mmWave frequency \cite{Wankai-mmWave_2021} bands, four-state digital surfaces for anomalous reflection in the sub-6 GHz and mmWave frequency bands \cite{Linglong-Access_2020}, varactor-controlled surfaces with continuous phase shift capabilities \cite{Romain-ContinuousTestbed_2021}, refracting surfaces in the mmWave frequency band \cite{Kyle-mmWallTestbed_2021}, and omni-coverage surfaces that are endowed with joint reflection and refraction capabilities \cite{Hongliang-OmniSurfaceTestbed_2021}. By contrast, a limited number of real-world measurement campaigns have been publicly reported. These include the demonstration of an RIS testbed that was deployed by Docomo and Metawave in a real urban environment in Japan for expanding the coverage of outdoor non-line-of-sight links at 28 GHz \cite{Docomo-OutdoorMeasurements_2019}, and a one-bit digitally-controlled RIS prototype that was successfully tested for data transmission over a 500 m communication link \cite{HaifanRIS-OutdoorMeasurements_2021}.

The available testbed prototypes and the field trials reported to date constitute fundamental proof-of-concept evaluations of the potential gains of RISs in realistic but small-scale settings. In order to evaluate the potential benefits of deploying RISs in existing and future wireless networks, large-scale simulations and field trials in realistic propagation environments and network topologies are needed. This article is a first attempt to shed light on the potential performance gains of RISs when they are deployed in a large-scale urban scenario and when they are integrated into an existing 5G cellular network. The simulation study reported in this paper is the first of its kind and relies on the Coffee Grinder Simulator\textsuperscript{\textcopyright}, which is a proprietary system-level platform for fast wireless system exploration that is developed and maintained by Huawei Technologies Sweden -- Gothenburg Research Center. The aim of our study is to quantify the performance gains offered by RISs, as a function of their size, deployment density, and frequency of operation, when they are deployed in an existing 5G cellular network. Specifically, we analyze and compare the deployment of RISs in the mid frequency band (the C-band at 3.5 GHz) and in the low mmWave frequency band (at 28 GHz). Several numerical results are illustrated and commented. For example, it is shown that the deployment of RISs enhances the coverage of cell-edge users from 77\% to 95\% at 3.5 GHz and from 46\% to 95\% at 28 GHz.

\begin{figure*}[!t]
	\centering
	\includegraphics[width=\linewidth]{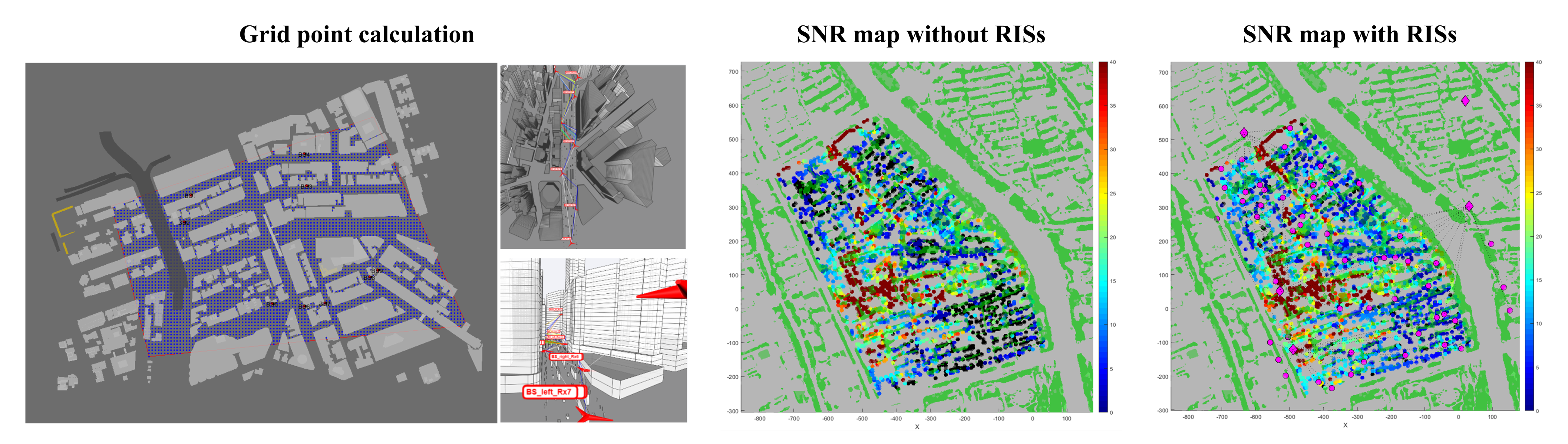}
	\caption{Coffee Grinder Simulator\textsuperscript{\textcopyright}: (left) grid-point calculation for the user locations, (center) SNR map without deploying RISs, and (right) SNR map with RISs deployed. The locations of the RISs are denoted with pink circles and the locations of the BSs are denoted with pink diamonds.}
	\label{Fig_Simulator}
\end{figure*}
\section{System-Level Simulation Platform}
To quantify the performance gains offered by the deployment of RISs in a typical urban environment that is served by a 5G network infrastructure, we employ the Coffee Grinder Simulator\textsuperscript{\textcopyright}. Coffee Grinder is a proprietary simulation platform for fast wireless systems exploration that is developed and maintained by Huawei Technologies Sweden -- Gothenburg Research Center. Coffee Grinder is an advanced link-level and system-level simulator that is designed for analyzing low and high frequency bands. The simulation platform is characterized by the agility of configuration in space and runtime, as well as the visualization of results. 

Coffee Grinder provides realistic urban and rural 5G and beyond 5G network deployments. The simulator integrates the OpenStreetMap API for modeling buildings, streets, vehicles, and it supports map-based 5G channel models based on the METIS 5G channel model, which includes vegetation and atmospheric attenuation models. Hybrid three-dimensional ray tracing models are supported based on the sparse point cloud method, i.e., a deterministic map-based model is employed for modeling surfaces and corners, and environmental small scale features are included through a diffuse scattering model. 

The propagation channels are calculated at discrete frequency bins and all components of the channel model are computed for all frequency bins, in order to obtain realistic and accurate results that account for the frequency dependency (e.g., antenna elements, beam squint, material properties, Doppler shift and spread, Fresnel zones, rain attenuation). The ray bundles are explored for all candidate routes and each ray is evaluated by applying Jones's calculus to the antenna, reflection, diffraction, penetration, and scattering. Several material models are supported, including those in the METIS channel model as well as custom materials. The antenna models of the base station (BS) and user equipment (UE) are fully customizable. High-level physical layer algorithms are supported as well. 

In Fig.~\ref{Fig_Simulator}, as an example, we illustrate the grid-based representation of the potential positions of the UEs in an urban environment, as well as an example of the ray bundles that can account for thousands of paths. As far as this article is concerned, the most important feature of the Coffee Grinder Simulator\textsuperscript{\textcopyright} is the support of realistic models for RISs, which include anaomalous reflections and the calculation of their locations for an optimized deployment. Coffee Grinder was initially designed in Matlab and it was subsequently migrated to Python with a GPU acceleration support (Nvidia Optics).

\begin{table*}[!t]
	\caption{Simulation setup}
	\label{Table_Setup}
	\centering
	\begin{tabular}{|c|c|c|}
		\hline 
		                                                      & \textbf{C-Band}                                     & \textbf{mmWave Band}                             \\ \hline 	                                         
		\textbf{City / Area (m\textsuperscript{2})}           & \multicolumn{2}{c|}{Urban Hangzhou, China / 800 x 800   m\textsuperscript{2}}                               \\ \hline	   
		\textbf{Frequency      }                              & 3.5 GHz                                              & 28 GHz                                            \\ \hline	   
		\textbf{Bandwidth (DL) }                              & 200 MHz                                              & 400 MHz                                           \\ \hline	   
		\textbf{\# BS       }                                 & 1                                                   & 5                                                \\ \hline	   
		\textbf{BS Height (m)  }                              & 30 m                                                 & 24-54 m                                           \\ \hline	   
		\textbf{BS Transmit Power   }                         & 50 dBm                                               & 32 dBm                                            \\ \hline	   
		\textbf{BS Array Gain  }                              & 23 dBi                                               & 30 dBi                                            \\ \hline	   
		\textbf{UE Array Gain   }                             & 3 dBi                                                & 10 dBi                                            \\ \hline	   
		\textbf{UE Placement  }                               & Indoor and Outdoor                                  & Outdoor                                          \\ \hline	   
		\textbf{3D UE Measurement   Grid   }                  & \multicolumn{2}{c|}{1 UE / 100 m\textsuperscript{2} (Ground/Floor)}                                     \\ \hline	   
		\textbf{UE Noise Figure   }                           & 7 dB                                                 & 10 dB                                             \\ \hline	   
		\textbf{Thermal Noise       }                         & -91 dBm                                              & -88 dBm                                           \\ \hline	   
		\textbf{RIS Width (m)   (square shape)}              & 2.70 m / 3.80 m / 5.30 m                               & 0.33 m / 0.48 m / 0.67 m                            \\ \hline	    
		\textbf{RIS Reflection   Losses }                     & 0 dB                                                 & 0 dB                                                    \\ \hline	   
		\textbf{\# Candidate RIS 3D   Positions }             & 4,000                                                & 34,000                                            \\ \hline	   
		\textbf{Channel Models}                               & \multicolumn{2}{c|}{
																\begin{tabular}{@{}c@{}}Metis2020 Propagation Model, \\ 
																						COST235 Vegetation Model, \\
																						Atmospheric Attenuation
																\end{tabular} 			
																		}                                                                                              \\ \hline	   
		\textbf{Cell-edge Target SNR  (SNR Coverage Target)} & \multicolumn{2}{c|}{10 dB (95\%)}  																	   \\ \hline	   
	\end{tabular}
\end{table*}
\section{Simulation Setup and RIS Modeling}
In this section, we describe the network simulation environment, the simulation setup, and how the RISs are modeled and incorporated into the system-level simulator.

In Table~\ref{Table_Setup}, we report the simulation setup. The considered network scenario corresponds to an urban area in the city of Hangzhou, China, which is identified by the geographical coordinates (30° 16' 58.8" N, 120° 9' 21.6" E). The area is covered by BSs that are located according to a real life grid. The number of BSs and RISs depends on the operating frequency. In this study, two typical carrier frequencies for 5G networks are analyzed: the C-band at 3.5 GHz (mid frequency band) and the mmWave band at 28 GHz (high frequency band). The focus of our analysis is on downlink (DL) transmissions. A similar analysis for the uplink (UL) can be performed. It is worth mentioning that the DL-UL reciprocity may not necessarily hold in RIS-aided channels, and it depends on the specific characteristics of the RISs \cite{Wankai-Sub6_2021}. The impact of the DL-UL reciprocity in RIS-aided networks is postponed to a further research work.

To analyze the improvement in terms of coverage and rate, we compute coverage maps across the entire considered geographic area, where the potential UEs may be located in indoors and outdoors. The UEs are assumed to be located on a grid-based layout with a density of 1 UE per 100 m\textsuperscript{2}. As far as the locations of the RISs are concerned, we have considered thousands of different three-dimensional candidate locations that are tested for improving the outdoor and indoor coverage. We assume that multiple RISs can assist the transmission of each UE. To better analyze the gains offered by deploying RISs, the BSs are assumed not to be equipped with massive MIMO antennas. The exact array gains of the BSs and UEs that are considered for link budget calculations are reported in Table~\ref{Table_Setup}. The use of massive MIMO antennas is expected to further improve the coverage and rate. Specifically, a UE is typically considered to be in coverage if the received SNR is greater than 0 dB. Besides this typical case study, we consider an SNR-boosted scenario in which the UEs are in coverage if the SNR is greater than 10 dB for each considered frequency bandwidth.

As far as the design of the RISs is concerned, we consider a typical far-field assumption. Therefore, we assume that the RIS is configured as an anomalous reflector and that it introduces a linear phase shift that depends on the difference between the angle of incidence and the desired angle of reflection. This is the typical geometrical optics approximation \cite[Eq. (94)]{PIEEE-CommunModels_2021} with unit amplitude constraint. It is known that the configuration of an RIS as an anomalous mirror is suboptimal in the near-field region of the RIS. However, it offers asymptotically the same performance as the optimal design (focusing lens) in the far-field region of the RIS \cite{Fadil-TCOM_2021}. From the practical point of view, deploying RISs that are configured as anomalous mirrors is advantageous since it is easier to estimate the directions of the UEs as compared with their exact locations. The accuracy of the far-field approximation is further discussed in the sequel (see Fig.~\ref{Fig_Fresnel}). The RISs are, in addition, designed according to a local design and no average power conservation is enforced, which usually results in local power amplifications. These assumptions result in a sort of ``worst case'' design for the RISs, which allows us to better understand the fundamental gains that the RISs can offer in a realistic 5G cellular network deployment.

As an illustrative example, Fig.~\ref{Fig_Simulator} shows the SNR coverage maps that are obtained with the Coffee Grinder Simulator\textsuperscript{\textcopyright} when the carrier frequency is 28 GHz. The maps are obtained by assuming that the SNR target for the cell-edge (5\textsuperscript{th} percentile) UEs is 10 dB. We see that 46\% of the UEs is in coverage (or 77\% if the SNR target is only 0 dB) in the absence of RISs, while 95\% of the UEs is in coverage when all the RISs in the network are utilized. It is apparent, therefore, that the deployment of RISs is beneficial for boosting the SNR of weak coverage areas and blind spots, which in turn enhances the coverage probability and the network spectral efficiency.

\subsection{RIS Deployment Algorithm}
The Coffee Grinder Simulator\textsuperscript{\textcopyright} offers the provision of deploying a large number of RISs throughout the considered geographical area. In addition, it offers the possibility of optimizing their locations given those of the UEs. Specially, the candidate locations for the RISs are assumed to be in line-of-sight with the available BSs and to form a finite grid. This grid includes locations that correspond to walls, roofs, and corners. The algorithm that is utilized for optimizing the locations of the RISs is reported below.
\begin{algorithm}[H]
	\caption{RIS Deployment Algorithm}
	\begin{algorithmic}[1]
		\STATE \textbf{Input:} Area map and BS locations;
		\STATE \textbf{Set} the target SNR threshold for cell-edge UEs (e.g,. ${\rm{SNR_{threshold}}}=10$ dB);
		\STATE \textbf{While} the target SNR criterion is not satisfied, i.e., $\mathcal{P}\{{\rm{SNR}} \leq {\rm{SNR_{threshold}}} \}>0.05$ \textbf{Do}
		\STATE \hspace{0.5cm}\textbf{Calculate} the RIS location (greedy-based exhaustive search) to maximize the number of UEs with ${\rm{SNR}} \geq {\rm{SNR_{threshold}}}$;
		\STATE \hspace{0.5cm}\textbf{Add} the RIS at the identified location;
		\STATE \hspace{0.5cm}\textbf{Calculate} the SNR at all possible locations of the UEs;
		\STATE \textbf{End}
		\STATE \textbf{Output:} RIS deployment map.														
	\end{algorithmic}
	\label{RIS_Optimization}
\end{algorithm}

Specifically, the algorithm for optimizing the positions of the RISs works as follows. First, the target SNR threshold is set for the cell-edge UEs (e.g., a cell-edge UE is in coverage if the SNR is greater than 10 dB). The implemented algorithm is iterative, and one new RIS is added at each iteration of the algorithm until the target SNR threshold is fulfilled. The placement of the each RIS at each iteration is obtained by utilizing a brute-force method (exhaustive search): the best location for the RIS is obtained by evaluating all possible combinations that account for the location of the BSs and the candidate locations of the RISs on the considered grid, so that the largest number of UEs fulfill the target SNR threshold. The considered approach is time consuming, since a large number of computations are needed and all the propagation paths need to be evaluated. Based on a sequential ordering, however, the algorithm returns the locations of the RISs that improve the SNR coverage. Therefore, it serves the main purpose and objective of this paper. The development of computationally efficient network deployment algorithms for RISs is postponed to a future research work.

\section{System-Level Simulation Results}
In this section, we report the system-level simulation results obtained with the Coffee Grinder Simulator\textsuperscript{\textcopyright} in the RIS-assisted 5G urban network scenario illustrated in Fig.~\ref{Fig_Simulator}. The system-level performance evaluation is conducted in terms of outdoor/indoor coverage and ergodic rate at different carrier frequencies and for different sizes of the RISs.

Specifically, the key performance indicator of interest is the cumulative distribution function (CDF) of the SNR received by the UEs in the absence and in the presence of RISs, i.e., the outage probability. Specifically, the complementary CDF (CCDF) represents the fraction (or percentage) of locations of the UEs where the received SNR is greater than the predefined target SNR threshold, i.e., the UE at the specified location is in coverage. From the obtained curves of CDF and CCDF, the performance gains of deploying RISs are quantified. The following two sections report the results for a 5G network deployment in the C-band and in the mmWave frequency range, respectively. Special focus is put on the impact of the size of the RISs at different carrier frequencies.

\begin{figure}[!t]
	\centering
	\includegraphics[width=\columnwidth]{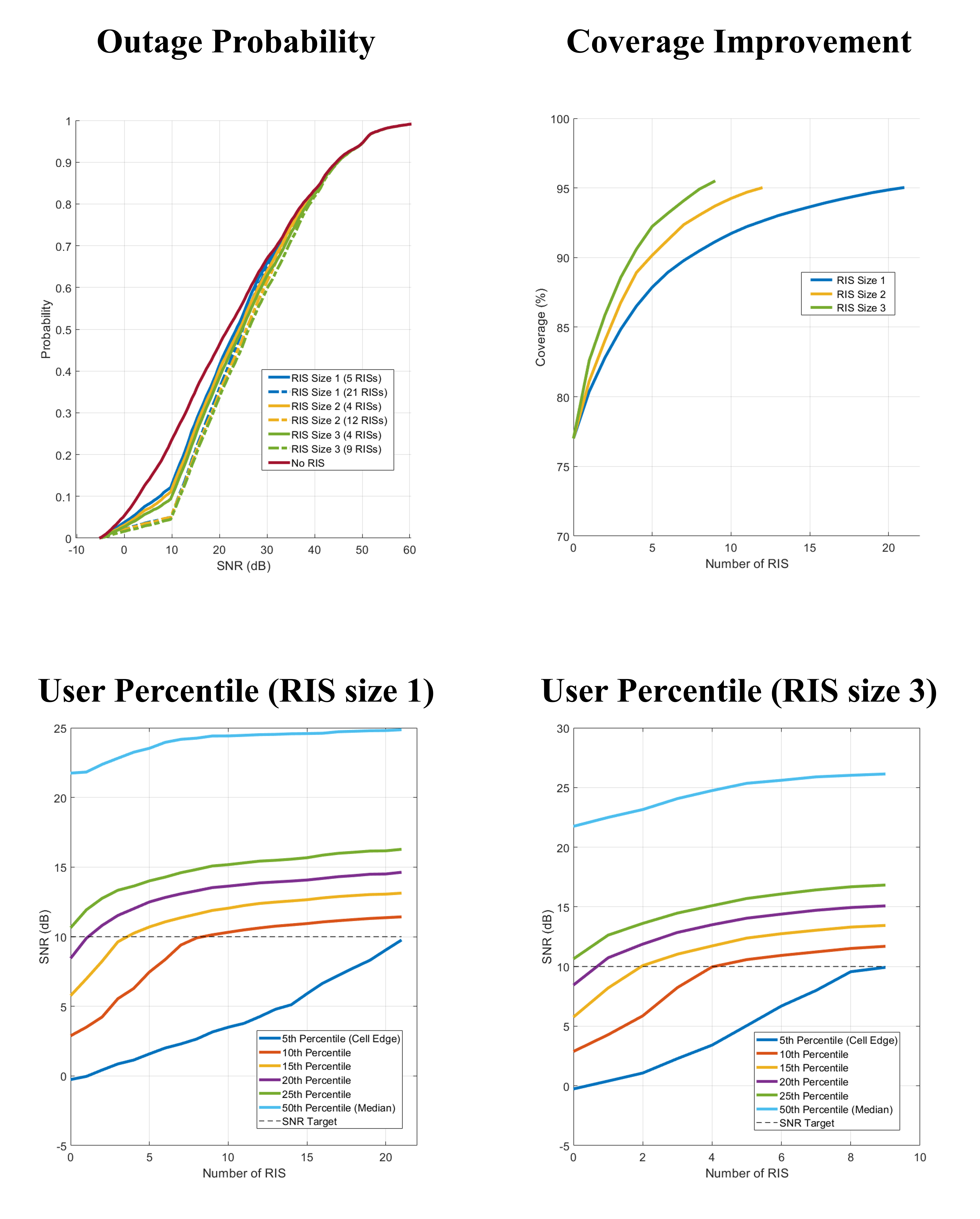}
	\caption{C-band (3.5 GHz) system-level analysis as a function of the size of the RISs. Size 1: 2.7 m x 2.7 m; size 3: 5.3 m x 5.3 m. (top) SNR outage and coverage of the 5\textsuperscript{th} percentile (cell-edge) of UEs when the target SNR threshold is 10 dB. (bottom) Received SNR for different percentiles of UEs as a function of the size of the RISs.}
	\label{Fig_Cband}
\end{figure}
\subsection{Case Study: C-band at 3.5 GHz}
In Fig.~\ref{Fig_Cband}, we report the system-level results for the 5G network deployment at 3.5 GHz. The outage probability and the SNR coverage are illustrated as a function of the size and the number of available RISs. Also, the received SNR of various percentiles of UEs is reported, including the median (often referred to as the typical) UEs (50\textsuperscript{th} percentile) and the cell-edge UEs (5\textsuperscript{th} percentile). The 5G network scenario encompasses one BS and several square-shaped RISs having three possible sizes: 2.7 m x 2.7 m (size 1), 3.8 m x 3.8 m (size 2), 5.3 m x 5.3 m (size 3). 

The two illustrations at the top of Fig.~\ref{Fig_Cband} show the SNR outage probability and the SNR coverage improvement of the 5\textsuperscript{th} percentile (cell-edge) of UEs when the target SNR threshold is 10 dB, which is a good SNR value for the cell-edge UEs. In the absence of RISs, the SNR coverage for the considered SNR target threshold is 77\%. A suitable target in RIS-aided networks is to attain 95\% coverage for the same target SNR  threshold of 10 dB and by considering both indoor and outdoor cell-edge UEs. The figure shows that this target is achieved if 21, 12, and 9 RISs of size 1, size 2 and size 3 are deployed in the network, respectively. As expected, the smaller the size of the RISs is the larger the number of surfaces to deploy is. In the considered case study of 3.5 GHz, a good compromise is obtained by using RISs whose size is 3.8 m x 3.8 m (size 2). Notably, we evince that only 5 RISs are needed for the 90\% of UEs to be covered, even though the SNR target threshold is 10 dB.

The two illustrations at the bottom of Fig.~\ref{Fig_Cband} show the received SNR for different percentiles of UEs. We note that the deployment of RISs provides a significant increase of the received SNR. The SNR of the typical user (the 50\textsuperscript{th} of UEs) increases of 3.1 dB and 4.4 dB if the size of the RISs is 2.7 m x 2.7 m (size 1) and 5.3 m x 5.3 m (size 3), respectively. It is worth noting that the SNR curves are steeper for a small number of RISs, which shows that the deployment of just a few RISs results in a substantial performance gain.

\begin{figure}[!t]
	\centering
	\includegraphics[width=\columnwidth]{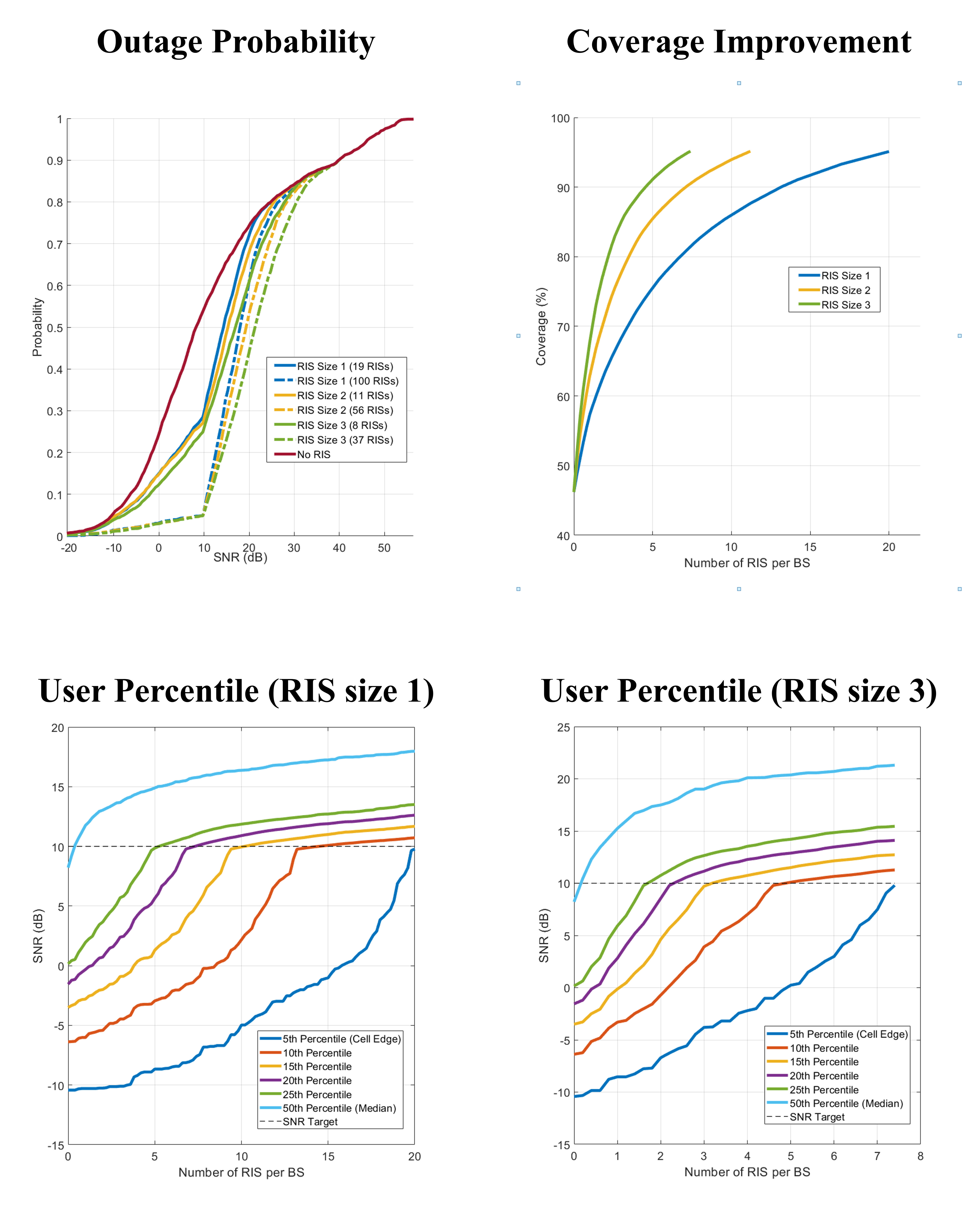}
	\caption{Millimeter-wave (28 GHz) system-level analysis as a function of the size of the RISs. Size 1: 0.33 m x 0.33 m; size 3: 0.67 m x 0.67 m. (top) SNR coverage of the 5\textsuperscript{th} percentile (cell-edge) of UEs when the target SNR threshold is 10 dB. (bottom) Received SNR of different percentiles of UEs as a function of the size of the RISs.}
	\label{Fig_mmWave}
\end{figure}
\subsection{Case Study: mmWave at 28 GHz}
In Fig.~\ref{Fig_mmWave}, we report the system-level results for the 5G network deployment at 28 GHz. The figure is tantamount to Fig.~\ref{Fig_Cband}, but the size of the RISs is chosen so as to ensure a similar number of tunable elements (unit cells) as for the case study in the C-band. More precisely, the sizes of the RISs are 0.33 m x 0.33 m (size 1), 0.48 m x 0.48 m (size 2), and 0.67 m x 0.67 m (size 3). As summarized in Table \ref{Table_Setup}, a main difference between the case study at 3.5 GHz and the case study at 28 GHz is that the number of BSs is increased from one to five. The results shown in  Fig.~\ref{Fig_mmWave} are similar to those in Fig.~\ref{Fig_Cband}, but there are some differences, qualitatively and quantitatively. First, we note that the coverage probability in the absence of RISs is only 46\%, even though only the outdoor UEs are considered in light of the high penetration losses in the mmWave band. The use of RISs that operate as refracting surfaces may overcome this limitation, as envisioned in some recent uses cases put forth by Docomo (see footnotes 2 and 3). The analysis of this deployment scenario is postponed to a future study.

From the two illustrations at the top of Fig.~\ref{Fig_mmWave}, we see that, with the aid of RISs, the coverage probability increases from 46\% to 95\% (still ensuring a target SNR threshold of 10 dB) by deploying, on overage, 20, 12, and 8 RISs of size 1, size 2, and size 3, respectively, per BS, which corresponds to a deployment of 100, 56, and 37 RISs in the considered geographic region. At 28 GHz, we evince, therefore, that RISs whose size is 0.48 m x 0.48 m (size 2) are a good compromise for ensuring that 95\% of the cell-edge UEs are in coverage by deploying 10 RISs per BS. Furthermore, the deployment of only 2 RISs per BS guarantees that more than 70\% (size 2) and 75\% (size 3) of the cell-edge UEs are in coverage.

From the two illustrations at the bottom of Fig.~\ref{Fig_mmWave}, we evince the impact of deploying RISs in the mmWave frequency band for different user percentiles. For example, the received SNR of the 5\textsuperscript{th} percentile of UEs (cell-edge UEs) increases by 20 dB if, for each BS, 20 RISs of size 1 or 8 RISs of size 3 are deployed. We note a large improvement of the SNR even for the median UEs (the 50\textsuperscript{th} percentile), since the SNR increases of the order of 9.8 dB (size 1) and 13.1 dB (size 3). Furthermore, we note that the slope of the curves is steep when the first RISs are added into the network, which implies that large SNR gains can be obtained even if a few RISs per BS are deployed.

\begin{table*}[!t]
	\caption{Coverage and ergodic rate by deploying RISs of different size in the C-band and mmWave band.}
	\label{Table_Gain}
	\centering
	\begin{tabular}{c|c|c c|c}
		\hline 
											           	 & \multicolumn{4}{c}{\textbf{Achievable Coverage [Required Number of RISs per BS]}}																 \\ \hline 
												         & \textbf{No RIS} 					& \multicolumn{2}{c}{\textbf{Few RISs}}				              & \textbf{Total RISs}              \\ \hline
		\textbf{RIS -- Size 1 at 3.5 GHz }		 & 77\% [0] 						& 80\% [1] 						   & 88\% [5] 						  & 95\% [21]                        \\ %\hline
		\textbf{RIS -- Size 2 at   3.5 GHz }      & 77\% [0]                         & 81\% [1]                         & 90\% [5]                         & 95\% [12]                        \\ %\hline
		\textbf{RIS -- Size 3 at   3.5 GHz }      & 77\% [0]                         & 83\% [1]                         & 92\% [5]                         & 95\% [9]                          \\ \hline %\hline
		\textbf{RIS -- Size 1 at   28 GHz  }      & 46\% [0]                         & 57\% [1]                         & 76\% [5]                         & 95\% [20]                        \\ %\hline
		\textbf{RIS -- Size 2 at 28 GHz    }      & 46\% [0]                         & 63\% [1]                         & 86\% [5]                         & 95\% [11.2]                      \\ %\hline
		\textbf{RIS -- Size 3 at   28 GHz  }      & 46\% [0]                         & 68\% [1]                         & 91\% [5]                         & 95\% [7.4]                       \\ \hline \hline
		
														         & \multicolumn{4}{c}{\textbf{Rate Improvement [Required Number of RISs per BS]}	}																   \\ \hline 
												         & \multicolumn{3}{c|}{\textbf{Cell-Edge}}								  & \textbf{Cell-Average }				 		                               \\ \hline
												         & \multicolumn{2}{c|}{\textbf{Few RISs}	}							  & \textbf{Total RISs}    				 & \textbf{Total RISs }                \\ \hline
		\textbf{RIS -- Size 1 at 3.5 GHz }		 & 3\% [1] 						 & 34\% [5] 						  & 253\% [21] 						 & 12\% [21] 						   \\ %\hline
		\textbf{RIS -- Size 2 at   3.5 GHz }      & 5\% [1]                         & 59\% [5]                         & 253\% [12]                         & 14\% [12]                         \\ %\hline
		\textbf{RIS -- Size 3 at   3.5 GHz }      & 11\% [1]                        & 115\% [5]                        & 258\% [9]                          & 16\% [9]                          \\ \hline
		\textbf{RIS -- Size 1 at   28 GHz  }      & 0\% [1]                         & 38\% [5]                         & 2508\% [20]                        & 45\% [20]                         \\ %\hline
		\textbf{RIS -- Size 2 at 28 GHz    }      & 8\% [1]                         & 238\% [5]                        & 2508\% [11.2]                      & 53\% [11.2]                       \\ %\hline
		\textbf{RIS -- Size 3 at   28 GHz  }      & 46\% [1]                        & 700\% [5]                        & 2508\% [7.4]                       & 62\% [7.4]                        \\ \hline
	\end{tabular}
\end{table*}
\subsection{Achievable Coverage and Rate Improvement}
Quantitatively, the main findings of the system-level simulation study are summarized in Table~\ref{Table_Gain}. Specifically, the table reports the achievable coverage and the improvement of the ergodic rate. As far as the coverage is concerned, the target SNR threshold is 10 dB and the number of RISs is obtained by assuming that 95\% of the UEs need to be in coverage. At both considered carrier frequencies, we evince that the deployment of a few RISs per BS results in a substantial increase of the coverage. The deployment of about 10 RISs guarantees that 95\% of the UEs have a received SNR equal to 10 dB. In the mmWave frequency band, even the deployment of a single RIS per BS results in increasing the coverage from about 46\% to about 65\%, depending on the size of the RISs. The deployment of only 5 RISs of appropriate size ensures that 90\% of the UEs is in coverage.

As far as the ergodic rate is concerned, in the absence of RISs the rate of the cell-edge UEs is only 0.96 bps/Hz at 3.5 GHz and 0.13 bps/Hz at 28 GHz. On the other hand, the ergodic rate that would correspond to a received SNR equal to 10 dB is, according to Shannon's formula, equal to 3.46 bps/Hz. Therefore, there is a substantial margin for improvement. Theoretically, in particular, the ergodic rate could be increased by 260\% and 2,561\% at 3.5 GHz and 28 GHz, respectively. From Table~\ref{Table_Gain}, we evince that the deployment of RISs with an appropriate size improves the ergodic rate by factors that are compatible with the theoretical optimum. If an appropriate number of RISs per BS is deployed, specifically, the ergodic rate is increased by a factor of 3.5 and by a factor of 25 in the C-band and in the mmWave band, respectively. It is worth noting that the deployment of RISs is not beneficial only for the cell-edge UEs, but the ergodic rate of the median (or typical) UEs is increased as well. In the absence of RISs, in fact, the average ergodic rate is 7.9 bps/Hz at 3.5 GHz and 4.8 bps/Hz at 28 GHz. In the mmWave frequency band, the ergodic rate of the typical UE is increased by a factor equal to 1.5.

Overall, the results reported in Table~\ref{Table_Gain} allow us to obtain insights on the tradeoffs that emerge for network planning in RIS-aided cellular networks as a function of the size and number of RISs per BS. These results can be taken into account along with the cost models for the RISs and BSs in order to identify the best deployment scenarios for RIS-aided networks.

\begin{figure}[!t]
	\centering
	\includegraphics[width=\columnwidth]{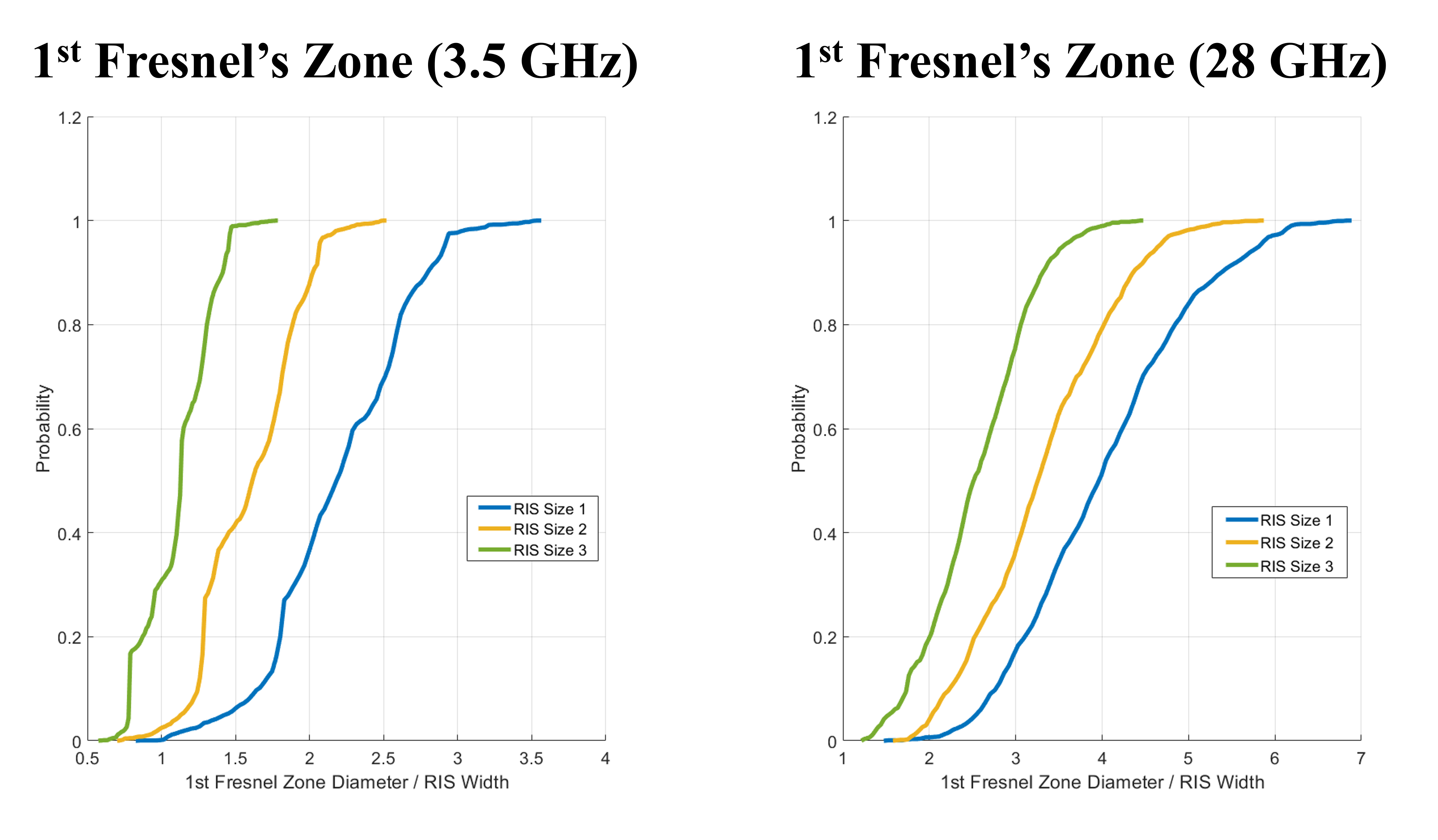}
	\caption{Near-field vs. far-field analysis at 3.5 GHz and 28 GHz: Distirbution of the ratio between the diameter of the first Fresnel zone and the width of the RISs.}
	\label{Fig_Fresnel}
\end{figure}
\subsection{On the Far-Field Assumption}
The system-level simulation results illustrated so far are obtained under the assumption that the RISs are configured as anomalous reflectors, and, therefore, they apply a linear phase modulation to the incident signals, which depends on the angle of incidence and the desired angle of reflection. As mentioned, this configuration of the RISs is suboptimal in the near-field region of the surface, while it is close to the optimum in the far-field region. Therefore, it is instructive to analyze the optimality of the obtained results from the point of view of utilizing RISs that are (sub-optimally) configured as simple anomalous reflectors. Specifically, the limiting distance at which an anomalous reflector is still a suitable option is the distance at which the RIS-aided path-loss is equal to the path-loss of the corresponding direct link \cite{Sibille-JSAC_2020}.

This analysis can be performed by resorting to the concept of Fresnel's zones \cite{Sibille-JSAC_2020}. By definition, in fact, the region of optimality of an RIS that is configured as an anomalous reflector is closely related to the diameter of the first Fresnel zone. If the ratio between the diameter of the first Fresnel zone and the width of the RIS is greater than one, then the RIS is sufficiently small for the configuration as an anomalous reflector to be nearly optimal. If the ratio is greater than one, on the other hand, the configuration of the RIS is not close to the optimum anymore and the RIS needs to be configured to operate as a focusing lens.

In Fig.~\ref{Fig_Fresnel}, motivated by these considerations, we analyze the probability distribution of the ratio between the diameter of the first Fresnel zone and the width of the RIS at 3.5 GHz and 28 GHz. In most cases, we observe that the ratio is greater than one and, therefore, the considered configuration for the RISs as anomalous reflectors is close to optimal. If the size of the RISs is 5.3 m x 5.3 m (size 3) in the C-band, we observe, however, that about 30\% of the UEs may be in positions for which the configuration of the RISs as anomalous reflectors is not optimal anymore. In this case, the RISs need to be optimized as focusing lenses. By taking into account that it may likely occur, or it may be added as an optimization constraint, that the optimal locations of the RISs provide a ratio between the diameter of the first Fresnel zone and the width of the RIS that is greater than one, we conclude that the use of RISs as anomalous reflectors may be a pragmatic choice for realistic network deployments, since such RISs are usually easier to realize and to optimize because it is simpler to estimate and track the angles of arrival of the radio waves with respect to the exact locations of the UEs.

\section{Closing Remarks}
RIS is an emerging technology for improving the coverage and the rate of wireless communication systems without requiring active power amplifiers, multiple radio frequency chains, and complex digital signal processing units. In this paper, for the first time in the open technical literature, we have reported a system-level performance evaluation study of an RIS-aided 5G network that is deployed in a typical urban environment for operation at 3.5 GHz (C-band) and 28 GHz (mmWave). The study is performed by utilizing the Coffee Grinder Simulator\textsuperscript{\textcopyright} that integrates a specifically designed algorithm for optimizing the locations of the RISs throughout the network, as a function of their size, operating frequency, and reflection capabilities. In particular, the RISs are modeled as anomalous reflectors for serving UEs located in the far-field region. 

The benefits of adding RISs have been evaluated in terms of coverage probability and ergodic rate for different user percentiles, which include the typical UEs (the 50\textsuperscript{th} percentile of UEs) and the cell-edge UEs (the 5\textsuperscript{th} percentile of UEs), by assuming that a UE is in coverage if the received SNR is greater than 10 dB in order to create coverage-boosted areas. Overall, the system-level simulations reveal that the deployment of RISs enhances the coverage of the cell-edge UEs from 77\% to 95\% in the C-band and from 46\% to 95\% in the mmWave band. The deployment of only 5 RISs per BS whose size is 3.8 m x 3.8 m in the C-band and 0.67 m x 0.67 m in the mmWave band ensures that 90\% of cell-edge UEs are in coverage. If 12 RISs per BS of the same size are deployed, the ergodic rate of the cell-edge UEs increases by a factor of 3.5 and by a factor of 25 in the C-band and mmWave band, respectively. In the mmWave band, in addition, the SNR of the cell-edge UEs increases by 20 dB if, for each BS, 20 RISs whose size is 0.33 m x 0.33 m or 8 RISs whose size is 0.67 m x 0.67 m are deployed. 

These results highlight the potential benefits of deploying RISs in a typical 5G urban network. Additional performance gains may be obtained by amalgamating RISs with other technologies, such as massive MIMO, and by considering more complex implementations of RISs that go beyond the operation as anomalous reflectors.

\vfill

\end{document}